\documentclass[twocolumn,showpacs,preprintnumbers,amsmath,amssymb]{revtex4}
\input psfig.sty

\usepackage{graphicx}
\usepackage{dcolumn}
\usepackage{bm}

\begin{document}

\title{Thermodynamics, spectral distribution and the nature of dark energy}

\author{J. A. S. Lima$^{1,2}${\footnote{limajas@astro.iag.usp.br}}
and J. S.
Alcaniz$^{2,3}${\footnote{alcaniz@dfte.ufrn.br}}}

\vspace{0.5cm}
\affiliation{$^{1}$Instituto de Astronomia, Geof\'{\i}sica e Ci\^encias
Atmosf\'ericas, USP, 05508-900 S\~ao Paulo - SP, Brasil\\
$^{2}$Departamento de F\'{\i}sica, Universidade Federal do Rio
Grande do Norte, 59072-970 Natal - RN, Brasil\\$^{3}$Observat\'orio Nacional, Rua
General Jos\'e Cristino 77, 20921-400, Rio de Janeiro - RJ, Brasil}

\date{\today}

\begin{abstract}
Recent astronomical observations suggest that the bulk of energy
in the Universe is repulsive and appears like a dark component
with negative pressure ($\omega \equiv p_x/\rho_x < 0$). In this
work we investigate thermodynamic and statistical properties
of such a component. It is found that its energy and temperature
grow during the evolution of the Universe since work is done on
the system. Under the hypothesis of a null chemical potential, the case of phantom
energy ($\omega < -1$) 
seems to be physically meaningless because its entropy is
negative. It is also proved that the wavelengths of the
$\omega$-quanta decrease in an expanding Universe. This unexpected
behavior explains how their energy may be continuously stored in the
course of expansion. The spectrum and the associated Wien-type
law favors a fermionic nature with $\omega$ naturally restricted
to the interval $-1 \leqslant \omega < -1/2$. Our analysis also
implies that the ultimate fate of the Universe may be considerably
modified. If a dark energy dominated Universe expands forever, it
will become increasingly hot.
\end{abstract}

\maketitle


The so-called dark energy or \emph{quintessence} is believed to be
the first observational evidence for new physics beyond the domain
of the Standard Model of Particle Physics. Its presence, inferred
from an impressive convergence of observational results along with
some apparently successful theoretical predictions, not only
explains the current cosmic acceleration but also provides the
remaining piece of information connecting the inflationary
flatness prediction with astronomical data \cite{prp}.

In spite of its fundamental importance for an actual understanding
of the evolution of the Universe, the nature of this unknown
energy component (which seemingly cannot be unveiled from
background tests \cite{padm}) constitutes one of the greatest
mysteries of modern Cosmology and nothing but the fact that it has
a negative pressure (and that its energy density is of the order
of the critical density, $\sim 10^{-29}$ g/cm$^{3}$) is known thus
far. This current state of affairs brings naturally to light some
important questions (among others) to be answered in the context
of this new Cosmology:

\begin{enumerate}
\item What is the thermodynamic behavior of the dark energy in an adiabatic expanding
Universe? Or, more precisely, what is its temperature law?
\item How do dark energy modes evolve in the course of the cosmic expansion?
\item What is the dark energy frequency spectrum?
\item What is the thermodynamic fate of a dark-energy-dominated
universe?
\end{enumerate}

In order to answer the above questions, it is necessary to make
some fundamental hypothesis concerning the unknown nature of the
dark energy. In principle, the thermodynamic behavior and the fate
of cosmic evolution are somewhat entertained and depend
fundamentally on the previous knowledge of the intrinsic nature of
the expanding components.

In this \emph{letter} we investigate the thermodynamic and some
statistical properties of the dark energy assuming that its
constituents are massless quanta (bosonic or fermionic) obeying a
constant equation-of-state parameter, $\omega = p_x/\rho_x$, where
$p_x$ is the pressure and $\rho_x$ its energy density. To be
more precise, most of the thermodynamic results remains true even
if the dark energy medium is formed by massive particles.

In what follows, answers to the above questions are sought by
using only the observational piece of information that for dark
energy $\omega$ is a negative quantity. Since $\omega$ can also be
seen as a continuous parameter, the case $\omega = 1/3$ (blackbody
radiation or neutrinos) provides a natural check for our general
results. Throughout this paper it will be assumed that the
Universe is described by the homogeneous and isotropic
Friedmann-Robertson-Walker (FRW) geometry
\begin{equation}
 ds^2 = c^{2}dt^2 - R^{2}(t) (\frac{dr^2}{1 - \kappa r^{2}}
  + r^2 d \theta^2 +
 r^2 sin^{2}\theta d\phi^2),
\end{equation}
where $\kappa = 0,\pm 1$ is the curvature parameter and $R(t)$
is the cosmological scale factor.

\emph{1. Dark Energy and Thermodynamics.} To begin discussing the
first question, we recall that the thermodynamic states of a
relativistic simple fluid are characterized by an energy momentum
tensor $T^{\alpha \beta}$, a particle current $N^{\alpha }$ and an
entropy current $S^{\alpha}$. By assuming that the dark energy is
a perfect fluid such quantities are defined and constrained by the
following relations \cite{dixon}
\begin{equation}
T^{\alpha \beta}=(\rho_x + p_x)u^{\alpha} u^{\beta} - p_xg^{\alpha \beta}, \quad
T^{\alpha \beta};_{\beta}=0 ,
\label{eq:TAB}
\end{equation}
\begin{equation} \label{eq:NA}
N^{\alpha}=nu^{\alpha}, \quad  N^{\alpha};_{\alpha}=0 ,
\end{equation}
\begin{equation} \label{eq:SA}
S^{\alpha}=n\sigma u^{\alpha}, \quad  S^{\alpha};_{\alpha}=0 ,
\end{equation}
where ($;$) means covariant derivative, $n$ is the particle number
density, $\sigma$ is the specific entropy (per particle) and the
quantities $p_x$, $\rho_x$, $n$ and $\sigma$ are related to the
temperature $T$ by the Gibbs law
\begin{equation} \label{eq:GIBBS}
nTd\sigma = d\rho_x - {\rho_x + p_x \over n}dn.
\end{equation}
By taking $n$ and $T$ as independent thermodynamic variables and
by using the fact that $d\sigma$ is an exact differential, it is
straightforward to show that the temperature evolution law is
given by \cite{weinb} (a dot means comoving time derivative)

\begin{equation} \label{eq:EVOLT}
{\dot T \over T} = \biggl({\partial p_x \over \partial
\rho_x}\biggr)_{_n} {\dot n \over n},
\end{equation}
or, equivalently, for $\omega \neq 0$
\begin{equation} \label{eq:TV}
n = \mbox{const}T^{1 \over  \omega}  \quad \Rightarrow \quad
T^{1/\omega}V = \mbox{const.},
\end{equation}
since $n$ scales with $V^{-1}$, where $V$ is the volume of the
considered portion within the fluid. In the case of blackbody
radiation ($\omega = 1/3$), the above equations yield the well
known results $n \varpropto T^{3}$ and $T^{3}V = \mbox{const.}$ It
means that blackbody radiation cools if it expands adiabatically,
a typical behavior for fluids with positive pressure ($\omega >
0$). Physically, this happens because each portion of the fluid is
doing thermodynamic work at the expenses of its internal energy.
>From Eq. (\ref{eq:TV}), however, we find that dark energy becomes
hotter in the course of the cosmological adiabatic expansion since
its equation-of-state parameter is a negative quantity. A physical
explanation for this behavior is that thermodynamic work is being
done on the system (see Fig. 1). In particular, for the vacuum
state ($\omega = -1$) we obtain $T \varpropto V$. Therefore, any
kind of dark energy with negative pressure (including the
so-called phantom energy, $\omega < -1$
\cite{phantom,alcaniz,pad}) becomes hotter if it undergoes an
adiabatic expansion. Naturally, this unexpected result needs to be
explained from a more fundamental (microscopic) viewpoint, which
is closely related to the remaining questions. However, before
discussing these questions, it is worth noticing that by combining
Eqs. (\ref{eq:TAB}) and (\ref{eq:TV}) one obtains the generalized
Stefan--Boltzmann law (see \cite{janilo} for some independent
derivations)
\begin{equation}\label{ROTETA}
\rho_x (T)=\eta_{\omega} T^{{1 + \omega \over \omega}},
\end{equation}
where $\eta_{\omega}$ is a $\omega$-dependent constant. For
$\omega=-1$ one finds $\rho_x =$ constant, as expected for the
cosmological constant case.  The entropy of this $\omega$-fluid is
also of interest. If the chemical potential is null (as occurs for
$\omega = 1/3$), the expression $\sigma = S_x/N = (\rho_x +
p_x)/nT$ defines its
 specific entropy \cite{degroot}. Therefore,
the dark energy entropy can be expressed as
\begin{equation}\label{ENTROP}
S_x(T,V) = \eta_{\omega}(1 + \omega)T^{1/\omega}V,
\end{equation}
from which the adiabatic temperature evolution law ($\ref{eq:TV}$)
is readily recovered. The above result completes our thermodynamic
description for the dark component. Note that the vacuum entropy
is zero ($\omega = -1$) whereas for phantom or
\emph{supernegative} dark energy ($\omega < -1$) the entropy
assumes negative values being, therefore, meaningless. If the
hypothesis of a null chemical potential is reasonable, this latter
result poses a serious problem to a phantom energy description
with basis on the usual XCDM parameterization and introduces a new
thermodynamic lower limit to the dark energy equation-of-state
parameter, i.e., $\omega \geq -1$. Such a limit is in fully
agreement with theoretical arguments that motivate the so-called
$\Lambda$ barrier\cite{alcaniz} and goes against several
observational analyses which seem to favor phantom cosmologies
over $\Lambda$CDM or usual quintessence
scenarios\cite{phantom,alcaniz,pad}. It should be recalled
that thermodynamic states with negative pressure are metastable
and usually connected with phase transitions (the most known
example at the level lab is an overheated Van der Waals
liquid \cite{Landau}). It should be stressed, however, that the
thermodynamic behavior derived here holds regardless of the
specific physical mechanism responsible for the negative
pressure.

\begin{figure}
\centerline{\psfig{figure=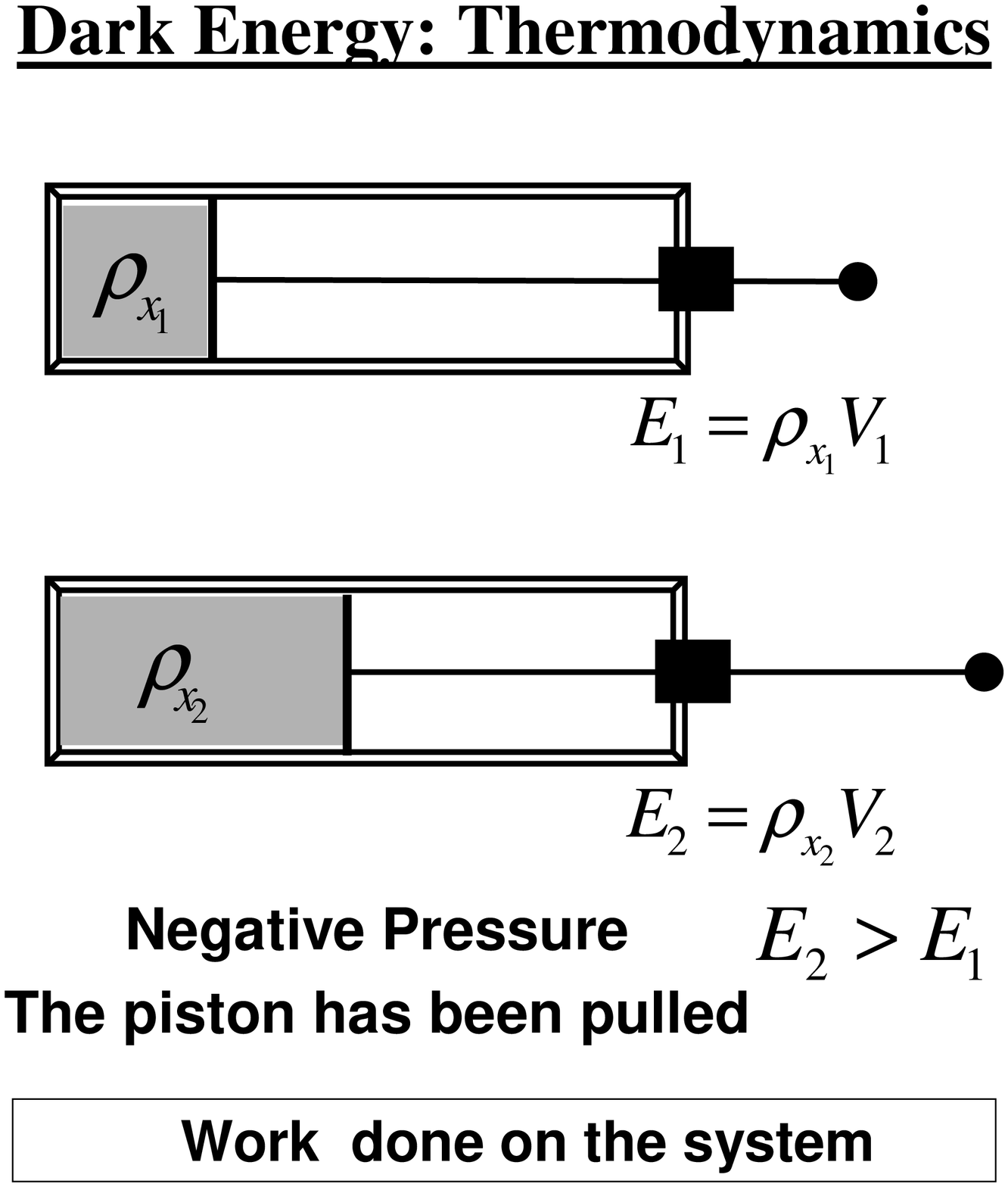,width=3.0truein,height=2.9truein}}
\caption{Dark Energy and thermodynamic work. The
total energy  and temperature  of the dark component grow in an
adiabatic expansion because work has been done on the system ($E_x
\equiv \rho_x V\varpropto T$). From Eqs. (7), (8) and (9) we see
that a vacuum filled Universe ($\omega = -1$) is a limiting case
where the temperature increases, the energy density remains
constant, and the entropy is zero.}
\end{figure}

\emph{2. Effects of the Expansion on the Dark Energy Modes.} In
what follows, we hipothesize that the constituents of the dark
energy are massless particles (bosons or fermions) and that the
differences between this \emph{fluid} and blackbody radiation (or
neutrinos) are macroscopically quantified by the equation-of-state
parameter $\omega$. As we shall see, this hypothesis leads to a
microscopic picture in harmony with the above thermodynamic
behavior.

The adiabatic theorem \cite{ehrenfest} guarantees that if a hollow
cavity (e.g., our Universe) containing dark energy changes adiabatically
its volume, the ratio between the energy of a given mode and the
corresponding frequency remains constant, that is, ${E_{\nu}/\nu
}=\mbox{const.}$, for any proper oscillation. Thus, if ${\rho_x}(T, \nu)$
is the spectral energy density inside an enclosure with volume $V$
at temperature $T$, this adiabatic invariant guarantees that
${{\rho_x}(T, \nu) d\nu V /\nu} = \mbox{const.}$ Now, recalling
that the energy density in the band $d\nu$ varies with the
temperature in the same manner as the total energy density, one
may write [see (\ref{ROTETA})]

\begin{equation}\label{TVNU}
{T^{\frac{1 + \omega}{\omega}}V /\nu} = \mbox{const.}
\end{equation}
and, since $T^{1 \over \omega}V=\mbox{const.}$, it follows that $T/\nu $ is
an invariant. Therefore, whether a hollow cavity containing dark
energy is expanding adiabatically, the wavelength of each mode
satisfies
\begin{equation}\label{LT}
\lambda T= \mbox{const.}
\end{equation}
The above results allow us to understand from a microscopic
viewpoint why the net energy of a dark component increases during
the expansion. As shown in Figure 2, the wavelengths of any
radiative fluid with positive pressure ($\omega > 0$) increase in
virtue of the expansion thereby lowering the total energy in
accordance with the generalized Stefan--Boltzmann law. However, if
$\omega < 0$, we have seen that the temperature grows whether the
fluid undergoes an adiabatic expansion [see Eq.(\ref{eq:TV})], and
this is accompanied by a decreasing in each wavelength $\lambda$
in order to satisfy the above relation. This, therefore, constitutes a microscopic
explanation of why the energy in a given portion of the dark
component increases (see Figure 1). In particular, the vacuum state, i.e., $\omega =
-1$, behaves as a limiting case for which the total energy
increases during the expansion while its energy density remains
constant.

\begin{figure}
\centerline{\psfig{figure=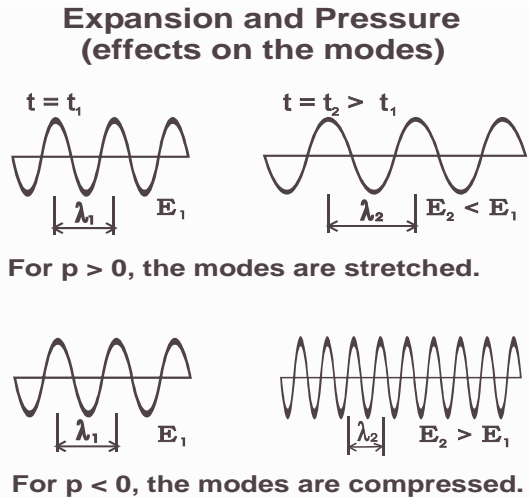,width=3.0truein,height=2.95truein}}
\caption{Pressure effect on the wavelength at two
different times ($t_{1}, t_{2}$) in an expanding Universe. The
wavelengths from radiation with positive (negative) pressure
increase (decrease) in the course of the expansion. Energy is
continuously stored in the dark energy modes ($p < 0$). This
unexpected behavior is a consequence of Eqs. (\ref{eq:TV}) and
(\ref{LT}). It explains why the internal energy of the dark
radiation increases during the evolution of the Universe.}
\end{figure}

\emph{3. The spectrum of dark energy.} Let us now quantify the
influence of a negative pressure on the general form of the
spectral distribution. This is an important point because
relations (\ref{eq:TV}) and (\ref{ROTETA}) must be recovered from
the frequency spectrum. To that end, we consider again an
enclosure containing dark energy at temperature $T_{1}$ and focus
our attention on the band $\Delta \lambda _{1}$ centered on the
wavelength $\lambda _{1}$ whose energy density is $\rho
_{x}(T_{1}, \lambda _{1})\Delta \lambda _{1}$. If the temperature
$T_{1}$ changes to $T_{2}$ due to an adiabatic expansion, the
energy of the band changes to $\rho _{x}(T_{2}, \lambda
_{2})\Delta \lambda _{2}$ and, according to Eq.(\ref{LT}), $\Delta
\lambda _{1}$ and $\Delta \lambda _{2}$ are related by ${\Delta
\lambda _{2}/\Delta \lambda _{1}} = {T_{1}/T_{2}}$. Now, since one
can assume that distinct bands do not interact, it follows that
\begin{equation}\label{RODELT}
{\rho _{x}(T_{2}, \lambda _{2})\Delta \lambda _{2} \over
\rho_{x}(T_{1}, \lambda _{1})\Delta \lambda _{1}}  = ({T_{2} \over
T_{1}})^{\omega + 1 \over \omega}.
\end{equation}
By combining the above results, and using again the adiabatic
invariant (\ref{LT}), we obtain for an arbitrary component
$\rho_{x} (T, \lambda ) = \alpha \, \lambda ^{-{1+ 2\omega \over
\omega}} \phi (\lambda T)$, or still, in terms of frequency
\begin{equation}\label{RONgN}
\rho_{x} (T, \nu ) = \alpha \nu ^{1 \over \omega}\phi ({\nu \over
T}),
\end{equation}
where $\alpha$ is a constant with dimension of
[$\rm{Energy}.\rm{Lenght}^{-3}.\rm{Time}^{(1 + \omega)/\omega}$],
and $\phi$ is an arbitrary dimensionless function of its argument.
The above expression is the generalized Wien-type spectrum for
dark energy.

The specific form of the dimensionless function $\phi (\nu/T)$ can
be determined using different approaches, among them, the
arguments put forward by Einstein in his original deduction of the
blackbody distribution \cite{einstein}. For a massless bosonic
gas, for instance,  $\phi (\nu /T)$ is exactly the occupation
number (without chemical potential), and the spectrum takes the
final form $\rho_{x}(T,\nu) = \alpha \nu^{1 \over \omega}
\times[e^{h\nu/k_{B}T}-1]^{-1}$ (see \cite{vac} for a more
detailed discussion). Therefore, in order to include the case of
fermions, we write the spectral distribution (13) as
\begin{equation} \label{eq:forend3}
\rho_{x}(T,\nu )={\alpha \nu ^{1\over \omega} \over e^{h\nu
/k_{B}T} \pm 1},
\end{equation}
which is the most natural extension of a Planck (or Fermi-Dirac)
spectrum describing the $\omega$-family of dark energy
\emph{radiation}. Einstein's derivation follows for $\omega
=1/3$ \cite{ajp} and, more important, the macroscopic relations
(\ref{eq:TV}) and (\ref{ROTETA}) are  recovered from the above
spectral distribution, i.e.,

\begin{equation} \label{eq:nCONST}
          n(T) = \int_{0}^{\infty} {\rho_x (T, \nu) \over h\nu}d\nu
          =\bar \eta_{\omega} \,  T^{{1 \over \omega}},
\end{equation}
and
\begin{equation} \label{eq:rOCONST}
         \rho_x (T) =\int_{0}^{\infty} \rho_{x}(T, \nu )d\nu =
         \eta_{\omega}\, T^{{1 + \omega} \over \omega},
\end{equation}
where the pair of constants ($\bar\eta_\omega, \eta_\omega$)
depend on $\alpha$, $k_B$ and $\omega$, and also on the underlying
statistics obeyed by the dark quanta.

Another interesting point is
related to the wavelength for which the distribution (\ref
{eq:forend3}) attains its maximum value. It is determined by the
algebraic condition
\begin{equation}
e^{-x} \pm 1 = \pm \omega (1 +
2\omega)^{-1}x,
\end{equation}
where $x={hc/k_B\lambda T}$ and the signs $\pm$
stand for the fermionic and bosonic cases, respectively. The
displacement Wien's law now reads
\begin{equation}\label{roots}
\lambda_m T = {hc \over k_{B}x'(\omega)} = {1.438 \over
x'(\omega)},
\end{equation}
where $x'(\omega)$ is the root of the above transcendent equation.
As one may check, in the case of bosons, a large set of nontrivial
solutions exist with positive $\omega$, but all of them are out of
the dark branch ($\omega < 0$). However, by assuming that the
basic constituents of dark energy are massless fermions, the
existence of solutions fixes the upper limit $\omega < -0.5$. In
particular, for $\omega = -2/3$, a value compatible with many
astrophysical data \cite{prp,adata}, we find $\lambda_m T=
1.943$ cm.K while for $\omega = -1$, $\lambda_m T = 1.123$ cm.K.

\emph{4. Dark energy and the thermodynamic fate of the expanding
Universe.} The fate of our Universe, i.e., whether it will
eventually re-collapse and end with a Big Crunch, or will expand
forever, is a matter of great interest among cosmologists (see,
e.g., \cite{cald}). In the standard view, if the Universe expands
forever it will become increasingly empty and cold. However,
whether our analysis provides a realistic description of the the
dominant dark energy component, such a destiny is not so neat
because of its strange thermodynamic behavior.

In the context of a FRW-type geometry ($V\propto R^{3}$), Eq.
(\ref{eq:TV}) implies that $T \varpropto R^{-3\omega}$ or,
equivalently, that the general redshift-temperature law reads
$T_{x}(z) = T^{o}_{x} (1 + z)^{3\omega}$, where $T^{o}_{x}$ is the
present value of the dark energy temperature. As discussed
earlier,  $T_{x}(z)$ grows in the course of the expansion while
the observed average temperature of the Universe is given by the
decreasing temperature of the CMB photons. Therefore, an important
point for the thermodynamic fate of the Universe is to know how
long the dark energy temperature will take to become the dominant
temperature of the Universe. A basic difficulty, however, is that
the present-day dark energy temperature has not been measured and,
moreover, the $\alpha$ parameter appearing in the spectrum [Eq.
(14)] (or, equivalently, the $\eta_{\omega}$ in the expression of
the energy density) is still unknown \footnote{Note that the
dimension of $\alpha$ is $h/c^{3}$ only for the standard case
($\omega = 1/3$). Possibly, for dark energy \emph{radiation}, such
a constant might be provided by a more fundamental theory.}. A
very naive estimate of $T^{o}_{x}$ can be done by considering the
observational evidence for the density parameters, $\Omega_x \sim
0.7$ and $\Omega_r \sim 10^{-4}$ (radiation). For $\omega \neq -1$
one finds $T^{o}_{x} \sim (10^{5} a/\eta_{\omega})^{\omega \over
{1 + \omega}}$, where $a$ is the radiation constant. Therefore,
any estimate of the present dark energy temperature depends
crucially on the ratio parameter $a/\eta_{\omega}$. In particular,
for $\omega = -2/3$ we find $T^{o}_{x} \sim
10^{-10}(\eta_{-2/3}/a)^{2}$, so that temperatures of the order of
$10^{-6}$K are easily obtained for $\eta_{-2/3}/a \sim 10^{2}$.
Albeit many aeons might be necessary for the Universe to enter in
the dark energy temperature regime, the only possible conclusion
is that the ultimate fate of the Universe may be considerably
modified: a dark energy dominated Universe expanding forever will
become increasingly hot.

Summarizing, we have proposed that the constituents of dark energy
are massless particles (bosons or fermions) whose collective
behavior resembles a kind of radiation fluid with negative
pressure. Through a thermodynamic analysis and basic statistical
considerations we derived some relevant properties obeyed by this
mysterious component. As expected for a consistent treatment, all
thermodynamic relations are recovered from the statistical
approach. Perhaps, more important, with basis on a Wien-type law
it was possible to show that a fermionic nature is clearly favored
because there are nontrivial solutions within the dark energy
branch ($\omega < 0$). It was also proved that the existence of
such solutions requires $\omega$ to be $< -0.5$. When combined
with the limit obtained from the entropy calculation [Eq. (9)],
this result constrains the equation-of-state parameter to the
interval $-1 \leq \omega < -0.5$, which is surprisingly close to
the range required from many different astrophysical data
\cite{prp,adata}. Such results means that some kind of
massless particles obeying the Fermi-Dirac statistics should be
considered among the candidates for dark energy proposed in the
literature. However, it should be stressed that our treatment does
not eliminate the case of massive fermions for which most of the
thermodynamic properties discussed here remains true.

\vspace{0.2cm}

{\bf Acknowledgments:} This work was partially supported by CNPq
(Brazilian Research Agency). We are also very grateful to N. 
Pires and R. C. Santos for helpful discussions.

\end{document}